# Residual stress induced stabilization of martensite phase and its effect on the magneto-structural transition in Mn rich Ni-Mn-In/Ga magnetic shape memory alloys

Sanjay Singh[1,2], Pallavi Kushwaha[2], F. Sceibel[1], Hanns-Peter Liermann[3], S. R. Barman[4], M. Acet[1], C. Felser[2] and Dhananjai Pandey[5]

[1]Experimentalphysik, University Duisburg-Essen, D-47048 Duisburg, Germany

[2] Max-Planck Institute for Chemical Physics of Solids, Nöthnitzer Strasse 40, 01187 Dresden, Germany

[3]Photon Sciences, FS-PE, Deutsches Elektronen Synchrotron (DESY), 22607 Hamburg, Germany

[4]UGC-DAE Consortium for Scientific Research, Khandwa Road, Indore, 452001, Madhya Pradesh, India.

[4]School of Materials Science and Technology, Indian Institute of Technology (Banaras Hindu University), Varanasi-221005, India.

## Abstract

The irreversibility of the martensite transition in magnetic shape memory alloys (MSMAs) with respect to external magnetic field is one of the biggest challenges that limits their application as giant caloric materials. This transition is a magneto-structural transition that is accompanied with a steep drop in magnetization (i.e., $\Delta M$) around the martensite start temperature ($M_s$) due to the lower magnetization of the martensite phase. In this communication, we show that $\Delta M$ around $M_s$ in Mn rich Ni-Mn based MSMAs gets suppressed by two orders of magnitude in crushed powders due to the stabilization of the martensite phase at temperatures well above the $M_s$ and the austenite finish ($A_f$) temperatures due to residual stresses. Analysis of the intensities and the FWHM of the x-ray powder diffraction patterns reveals stabilized martensite phase fractions as 97, 75 and 90% with corresponding residual microstrains as 5.4, 5.6 and 3% in crushed powders of the three different Mn rich Ni-Mn alloys, namely, $Mn_{1.8}Ni_{1.8}In_{0.4}$, $Mn_{1.75}Ni_{1.25}Ga$ and $Mn_{1.9}Ni_{1.1}Ga$, respectively. Even after annealing at 773 K, the residual stress stabilised martensite phase does not fully revert to the equilibrium cubic austenite phase as the magneto-structural transition is only partially restored with reduced value of $\Delta M$. Our results have very significant bearing on application of such alloys as inverse magnetocaloric and barocaloric materials.

Recent years have witnessed tremendous surge in the study of ferroic and multiferroic materials exhibiting giant caloric effects that can be used in solid state refrigeration at temperatures close to the ambient conditions as an environmentally friendly substitute to conventional vapour compression refrigeration [1-5]. The Heusler type Mn rich Ni-Mn-X (X = Ga, In, Sn, Sb) MSMAs have emerged as a potential family of alloys that can exhibit giant barocaloric, elastocaloric and inverse magnetocaloric (i.e., cooling during magnetization and heating during demagnetization in contrast to normal magnetocaloric materials which heat up on magnetisation and cool down by its removal) effects [6-10]. The giant inverse MCE and BCE in these alloys is linked with a first order martensitic transition, which is a magnetostructural transition involving change of crystal structure as well as huge $\Delta M$ between the austenite and martensite phases [11]. This transition involves large isothermal entropy change ($\Delta S_{iso}$) and hence large adiabatic temperature change ($\Delta T_{ad}$) that leads to the caloric effect [8, 12]. One of the major limitations of these otherwise potentially promising Heusler alloys is the irreversibility of the martensitic transition as a function of magnetic field cycles due to the sluggish kinetics of the first order structural phase transition [8, 13-15]. As the MSMAs owe their large inverse MCE and BCE due to strong coupling of magnetic and elastic degrees of freedom, the irreversibility of the $\Delta S_{iso}$ or the $\Delta T_{ad}$ with respect to magnetic field and pressure is expected to be closely related with the residual stresses and strains developed in these materials as a result of magnetic field and/or stress (pressure) cycles.

In the present work, we have investigated the effect of residual stresses on the phase stabilities and magnetization behaviour in two different class of Mn- rich Ni-Mn-X (X=Ga and In) MSMAs whose martensite start temperature $M_s$ is close to the room temperature while the ferromagnetic curie temperature (Tc) is greater than $M_s$. Using high resolution synchrotron and laboratory x-ray powder diffraction data, it is shown that the martensite structure can be stabilized by residual stresses over a wide temperature range well above $A_f$. As a direct consequence of the residual stress induced stabilization of the martensite phase, the peak value of magnetization ($M$) around $M_s$ and the associated change in $M$ (i.e., $\Delta M$) between the austenite and martensite phases is shown to decrease by two orders of magnitude. Even after annealing at 773 K, the original value of $\Delta M$ corresponding to the bulk sample is not fully recovered due to the presence of retained martensite. Our results on Ni-Mn-In/Ga alloys unambiguously demonstrate that the residual stresses strongly affect $\Delta M$. We believe that our findings will have very significant bearing on the application of Mn rich Ni-Mn type MSMAs as giant inverse magentocaloric and barocaloric materials.

The three alloy compositions investigated by us are: $Mn_{1.8}Ni_{1.8}In_{0.4}$, $Mn_{1.75}Ni_{1.25}Ga$ and $Mn_{1.9}Ni_{1.1}Ga$. The details of sample preparation, measurements (magnetization and x-ray powder diffraction (XRD)) and Rietveld refinements are given in the supplementary file. For the $Mn_{1.8}Ni_{1.8}In_{0.4}$ alloy, we have investigated the magnetization and structural characteristics for annealed pieces cut from the bulk ingot, powder samples obtained by crushing the ingot, and powder sample annealed under different conditions. The results for bulk ingot piece and crushed powder are also presented for the Ga alloys to demonstrate the likely universality of the phenomenon of the stabilization of the martensite phase due to residual stresses in Mn rich Ni-Mn MSMAs.

The magnetization as a function of temperature ($M$(T)) for annealed $Mn_{1.8}Ni_{1.8}In_{0.4}$ bulk sample (melt-ingot-annealed piece), as measured under zero field cooling (ZFC), field cooling (FC) and field warming (FW) conditions using a magnetic field of 100 Oe, reveals a ferromagnetic transition at $T_C \sim 316$ K (see Fig. 1). The $M_s$, martensite finish ($M_f$) and austenite start ($A_s$) temperatures related to the austenite to martensitic and martensite to austenite transitions during cooling and warming are found to be 309 K, 290 K and 296 K, respectively. The low field magnetization behavior as a function of temperature for the as-ground powder of $Mn_{1.8}Ni_{1.8}In_{0.4}$ is drastically different from that of the annealed sample. The $M$ (T) plots recorded on the as-ground powder during cooling under 100 Oe from 300 K to 2 K (C1), subsequent warming (at 100 Oe) up to 400 K (C2) and then cooling (at 100 Oe) down to 2 K (C3) are shown in Fig 2. It is evident from a comparison of this figure (Fig. 2) with Fig. 1 that during the first cooling (C1) and warming (C2) cycles, the peak value of the magnetization associated with the austenite-martensite transition is about two orders of magnitude lower than that of the annealed sample. In contrast to the C1 and C2 cycles, comparatively larger change in magnetization associated with the austenite-martensite transition is observed when the measurement was carried out during cooling (Fig 2, C3) the powder from 400 K. The anomalies in the magnetization plot at the Tc and $M_S$ transitions are similar to the annealed sample (Fig.1) but the peak value of the magnetization is still an order of magnitude lower than that for the annealed sample. Similarly an increase in the magnetization is also observed when powder was warmed from 2K to 400 K (C4) after the C3 cycle but the peak value remains an order of magnitude lower than the corresponding value for the annealed bulk sample. However, with further warming and cooling cycles after C4 on the same powder [Inset of Fig. 2 C5(ZFC), C6 (FC) and C7 (FW)], there was no change in the peak magnetization value at the austenite-to martensite and reverse transitions as the M(T) curve nearly coincides with those corresponding to the C3 and C4 cycles. Our results thus demonstrate that the external stresses introduced during crushing of the ingot into powder affect the magneto-structural transition drastically.

In order to understand the anomalously low values of peak magnetization of the as-ground powders as compared to that of the annealed bulk sample, we now proceed to present the details of structural studies on the as-ground and annealed samples. The powder XRD profiles of the 220 austenite peak for the as-ground and annealed $Mn_{1.8}Ni_{1.8}In_{0.4}$ powders shown in Figs.1 (a) and 1(b) of the supplementary file reveals a change in crystal structure after annealing. The XRD pattern for the as-ground powder of $Mn_{1.8}Ni_{1.8}In_{0.4}$ was recorded at T= 340 K, which is well above the $A_f$ (~317K (estimated from the slope of FW curve for the magnetization value similar to FC at $M_s$). At this temperature, the sample is expected to be in the stable austenite cubic phase (see Fig.1). The Rietveld analysis of the XRD patterns recorded at 340 K ($> M_s$) shows that it contains ~97% tetragonal martensite and ~ 3% cubic austenite phases in the I4/mmm and Fm-3m space groups, respectively, (see Fig 3(a) for the fit between observed and calculated profiles). The tetragonal martensite phase is not expected to exist at 340 K as this temperature is ~30 K higher than the $M_s$ ~ 309K and it is also higher than $A_f$. Rietveld refinement of the as-ground powder sample after annealing at 773 K confirms that the tetragonal phase has transformed to the thermodynamically stable cubic austenite phase (Fig. 3b). Evidently, the annealed sample reveals the thermodynamically stable crystal structure whereas the as-ground sample corresponds to a stress induced martensite phase. There is, however, a very small fraction (~3%, as estimated from the peak intensities of the most intense peaks of the two phases) of the untransformed cubic austenite phase whose presence can also be seen in the inset of Fig 3b. In case of conventional shape memory alloys, the stress induced martensite phase formed above $A_f$ temperature reverts back to the austenite phase on removal of the stress and this reversibility of the austenite-martensite transition is responsible for the pseudoelastic behavior observed in those alloys [16, 17]. In our case, the crushed sample continues to exhibit the stress induced martensite phase suggesting that the transformation is nearly irreversible in crushed powders and is therefore detrimental to caloric effects that generate stress directly through pressure [7] / compressive uniaxial stress [9] or indirectly through magnetic field [11].

It is interesting to note that the Bragg peaks of the as-ground powders are very broad and become sharper after annealing (see Fig.1 of the supplementary file that shows Synchrotron XRD data). The most likely sources of broadening in the as-ground samples are the residual stresses and the size of the different martensite variants. To separate out the two contributions to peak broadening, we carried out Williamson–Hall (W-H) analysis of the peak widths ($\beta$) of the various Bragg peaks. In the Williamson–Hall analysis, the average microstain ($\varepsilon = <\Delta d/d>$) and coherently scattering domain size ($<D>$) are determined from the relationship $\beta \cos\theta = 2<\Delta d/d>\sin\theta + \lambda/<D>$, where $\theta$ is the Bragg angle, $\lambda$ the wavelength of the x-rays and $\beta$ is the full width at half maximum (FWHM) of the peak after subtracting the FWHM of standard sample (which in our case was $LaB_6$). The W-H plot for the as-ground powder sample of $Mn_{1.8}Ni_{1.8}In_{0.4}$ shows two straight line fits to the data points corresponding to various reflections (see Fig.2 of the supplementary file) indicating anisotropic peak broadening with the largest slope giving a residual microstrain of ~ 5.4 %. The two intercepts on the ordinate axis give coherently scattering domain sizes of ~43 Å and 345 Å respectively, corresponding to the thin martensite variants with large lateral extension (such a thin plate like morphology of the martensite phase has been reported in Ref [18]). After annealing the powder at 773 K, the microstrain was reduced to 0.02% and the domain size increased to 550 Å. The presence of large microstrains in the as-ground powder of $Mn_{1.8}Ni_{1.8}In_{0.4}$ at 340 K clearly suggests that the martensite phase is stabilized by the residual stresses introduced during grinding of the ingot sample into powder. It is also interesting to note that a part of the anisotropic peak broadening is due to the change in the coherently scattering domain size which is smaller in the martensite phase due to the formation of submicron size martensite variants in thin plate like morphologies and larger in the cubic austenite phase as there are no such variants (see Table1 of the supplementary file for the values of microstrain, domain size and unit cell parameters).

The low peak values of magnetization at the martensite-austenite transition for the C1 and C2 cycles in Fig 2 is evidently linked with the stabilization of the martensite phase well above the $M_s$ and $A_f$ due to the residual stresses discussed above. In the absence of the austenite phase (fully stabilized martensite), the peak in M (T) around $M_s$ is not anticipated as the peak appears due to the difference in the magnetization of the austenite and martensite phases. A small peak that appears in M (T) during C1 and C2 cycles is due to a very small fraction (~3% as determined by the relative intensity of most intense peak of the cubic austenite phase i.e. 220) of the untransformed austenite phase. The increase in the magnetization value after taking the as-ground powder up to 400 K (C3 and C4) is because of the increase in the fraction of the austenite phase. However, the temperature of 400 K is not sufficient to transform the entire martensite phase and as a result the peak values of the magnetization at the austenite- martensite transition are still an order of magnitude lower than those for the annealed bulk samples. Even after annealing at much higher temperatures (773 K), the peak value for the magnetization (~ 3 emu/gm) at the martensite-austenite transition (Fig.4) still remains only ~ 50 % of the peak value for the bulk ingot. This is because the stabilized martensite phase is still present in very small fractions in the 773 K annealed powders as can be seen from the tiny peaks in the inset of Fig 1(b) of the supplementary files. Our results thus show that full recovery of the magnetostructural characteristics is not possible by annealing unless the sample is recrystallized after melting.

It is interesting to note that residual stresses enhance the $M_s$ and lower the $T_C$ (see inset of Fig. 4, which depicts the normalized derivative plot dM/dT versus T) similar to what has been reported in the studies related to barocaloric [7, 19] and elastocaloric [9] effects. The increase in the magneto-structural transition temperature $M_s$ due to the residual stresses is consistent with the Clausius-Clapeyron equation "dT/dp= $\Delta V/ \Delta S$" (here $\Delta S$ and $\Delta V$ are the entropy and volume changes, respectively at the phase transition) for a first order phase transition [7, 19]. In the present case, the residual stresses play the role of pressure which not only enhances the $M_s$ but also stabilize the martensite phase well above $M_s$ and $A_f$ in the powder samples due to its lower volume as compared to that of the austenite phase. The opposite behavior of $T_C$ indicates reduced ferromagnetic exchange interaction in the plastically deformed cubic austenite regions that did not transform due to deformation. Our results therefore have direct bearing on the giant barocaloric and elastocaloric behavior of such alloys as the pressure may not only stabilize the martensite phase well above the $M_s$ and $A_f$ temperatures and reduce the peak value of the bulk magnetization at the magnetostructural transition temperature but also make the transition partially irreversible. A similar irreversibility of the magnetostructural phase transition induced by magnetic field is also anticipated because of the internal strains generated by the magnetic field [11] and this may adversely affect the magnetocaloric properties as well.

The residual stress induced stabilization of the martensite phase is not limited to In based Ni-Mn MSMAs as we have observed similar effects in two other Mn excess Ni-Mn-Ga alloys: $Mn_{1.75}Ni_{1.25}Ga$ and $Mn_{1.9}Ni_{1.1}Ga$, which also show inverse magnetocaloric effect [10]. The $M$(T) plot in a low applied magnetic field of 50 Oe for annealed bulk samples shown in Fig. 5 (a) reveal $M_s$, $M_f$, As and $A_f$ for $Mn_{1.75}Ni_{1.25}Ga$ as 139 K, 134 K, 160 K and 175 K, respectively, while the corresponding temperatures for $Mn_{1.9}Ni_{1.1}Ga$ are 264 K, 160 K, 230 and 315 K. The drop at $M_s$ is related to the large magnetocrystalline anisotropy and lower magnetization of the martensite phase in these MSMAs [10, 20, and 21]. A comparison of x-ray powder diffraction profiles of the as- ground powder samples and the same powder after it was annealed at 773 K for 10 hrs. (Fig.5), as was done for $Mn_{1.8}Ni_{1.8}In_{0.4}$ (Fig.3), shows that the martensite phase has been stabilized in these alloys also at room temperature, even though the room temperature is considerably higher than $M_s$ as compared to the Ni-Mn-In alloy system, due to stresses introduced during crushing. Fig.5 (b) shows the Rietveld refinement results for the as-ground powder samples while Fig 5 (c) for the annealed powder samples. The Rietveld refinements reveal that the as-ground $Mn_{1.75}Ni_{1.25}Ga$ and $Mn_{1.9}Ni_{1.1}Ga$ have tetragonal structure (space group I4/mmm) with some retained cubic (space group Fm-3m) austenite phase that corresponds to ~ 25% and 10% for Mn$_{1.75}$Ni$_{1.25}$Ga and Mn$_{1.9}$Ni$_{1.1}$Ga, respectively (see Table.1 of the supplementary file for more details). The Rietveld refinements of the powder samples after annealing at 773 K confirm that the tetragonal phase (Fig.5 (b)) reverts to the thermodynamically stable cubic austenite phase after annealing (Fig.5 (c)). The W–H analysis shows large residual microstrain of about 5.6% and 3% in the as-ground powder samples of Mn$_{1.75}$Ni$_{1.25}$Ga and Mn$_{1.9}$Ni$_{1.1}$Ga, which are reduced to 0.5% and 0.35%, respectively, after annealing the powder at 773 K in the cubic austenite phase (Fig.5 (c)). A similar effect has also been observed for Mn$_2$NiGa [22], where the martensitic phase is stabilized at temperatures higher than M$_S$ and A$_f$.

In conclusion, we have shown that the Mn rich Ni-Mn based MSMA's are highly sensitive to residual stresses, which can stabilize the martensite phase far above the martensite transition temperature. The fact that this effect is observed for both In and Ga based Ni-Mn-X alloys shows that this is not related with the active element (X= In or Ga) and that it is the excess Mn that is crucial to the stabilization of the martensite phase well above the martensite start and austenite finish temperatures in these systems. This stabilization drastically affects the characteristics of the magneto-structural transition because of the lower magnetization of the martensite phase and will therefore have important bearing on applications of these alloys as magnetocaloric and barocaloric materials.

S. S. thanks Alexander von Humboldt foundation, Germany for Fellowship. DP acknowledges the financial support for the x-ray diffraction studies using synchrotron radiation under the DST-DESY project operated through Saha Institute of Nuclear Physics and Science and Engineering Research Board of India for financial support through the award of J C Bose National Fellowship.

**Figures:**

Fig.1: The variation of zero field cooled (ZFC), field cooled (FC) and field warming (FW) magnetization of $Mn_{1.8}Ni_{1.8}In_{0.4}$ with temperature measured at 100 Oe. Ms is the martensite start temperature.

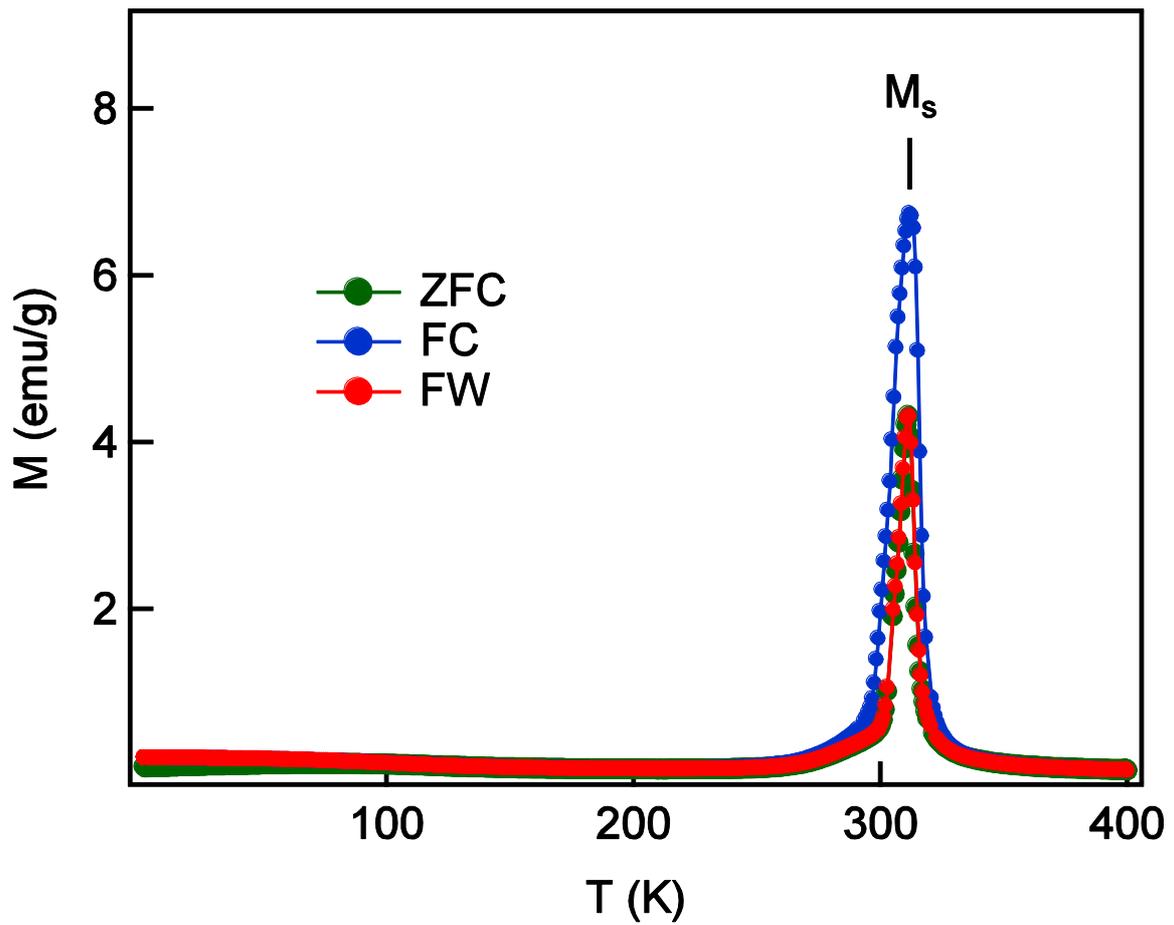

Fig.2: Magnetization as a function of temperature for as-ground powder sample of Mn$_{1.8}$Ni$_{1.8}$In$_{0.4}$ measured at H= 100 Oe (see text for description of C1, C2, C3, C4). Inset shows the successive data cycles taken after C4 (C5, C6, C7).

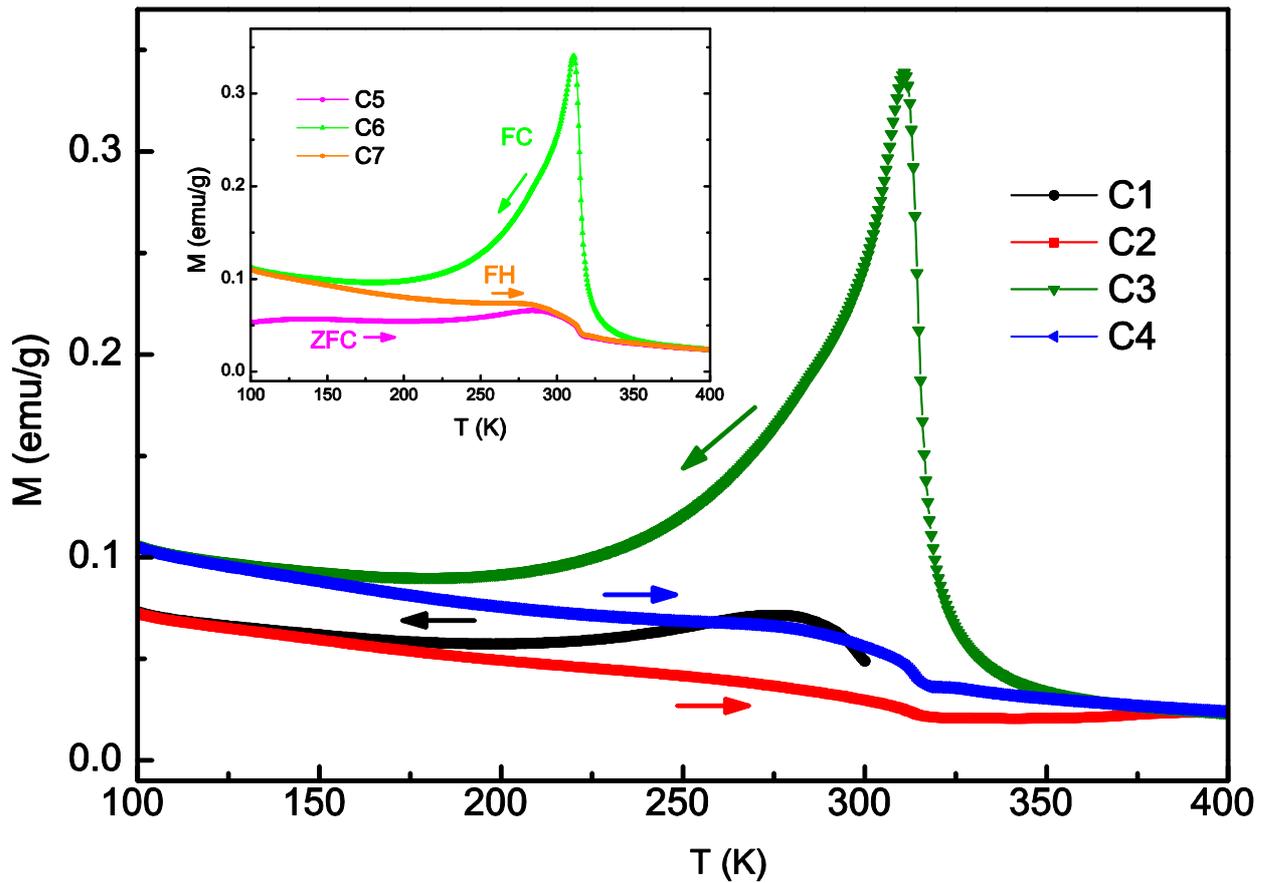

Fig.3: The observed (open circles), calculated (red solid line) and difference (green solid line) profiles obtained after Rietveld refinement of the structure of Mn$_{1.8}$Ni$_{1.8}$In$_{0.4}$ considering tetragonal (space group: I4/mmm) martensite and cubic austenite (space group: Fm-3m) phases: (a) as-ground powder sample and (b) after vacuum annealing the same powder at 773 K for 10 hrs. Inset in (a) shows the presence of austenite peak (marked as C). Blue ticks represent the Bragg peak positions.

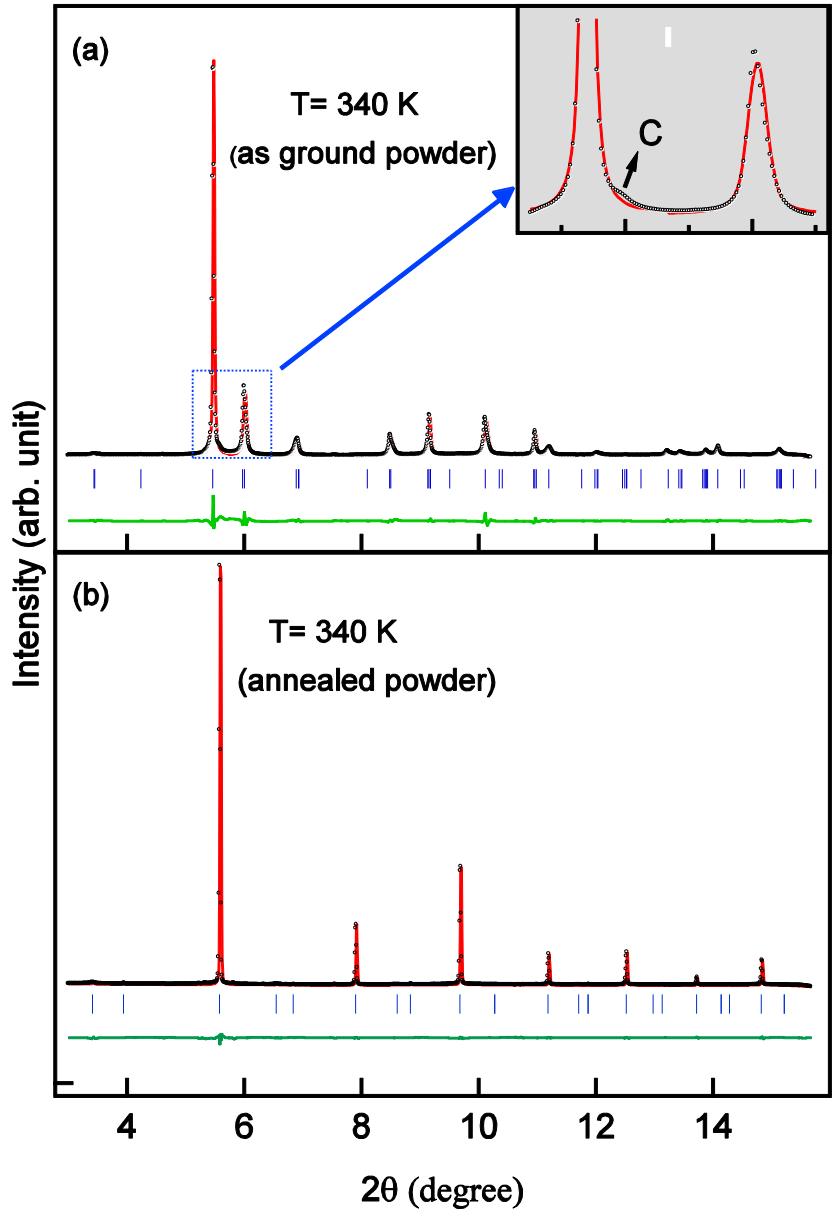

**Fig.4:** Comparison of field cooled magnetization (H= 100 Oe) as a function of temperature for annealed bulk, as-ground powder and annealed powder (at 773 K). Inset shows the normalized derivative of magnetization (dM/dT). The $M_s$ obtained from the derivative plot for annealed bulk, as-ground powder and annealed powder are ~306 K, 308 K and 309 K, while the Tc are ~314.6 K., 315 K and 316 K, respectively.

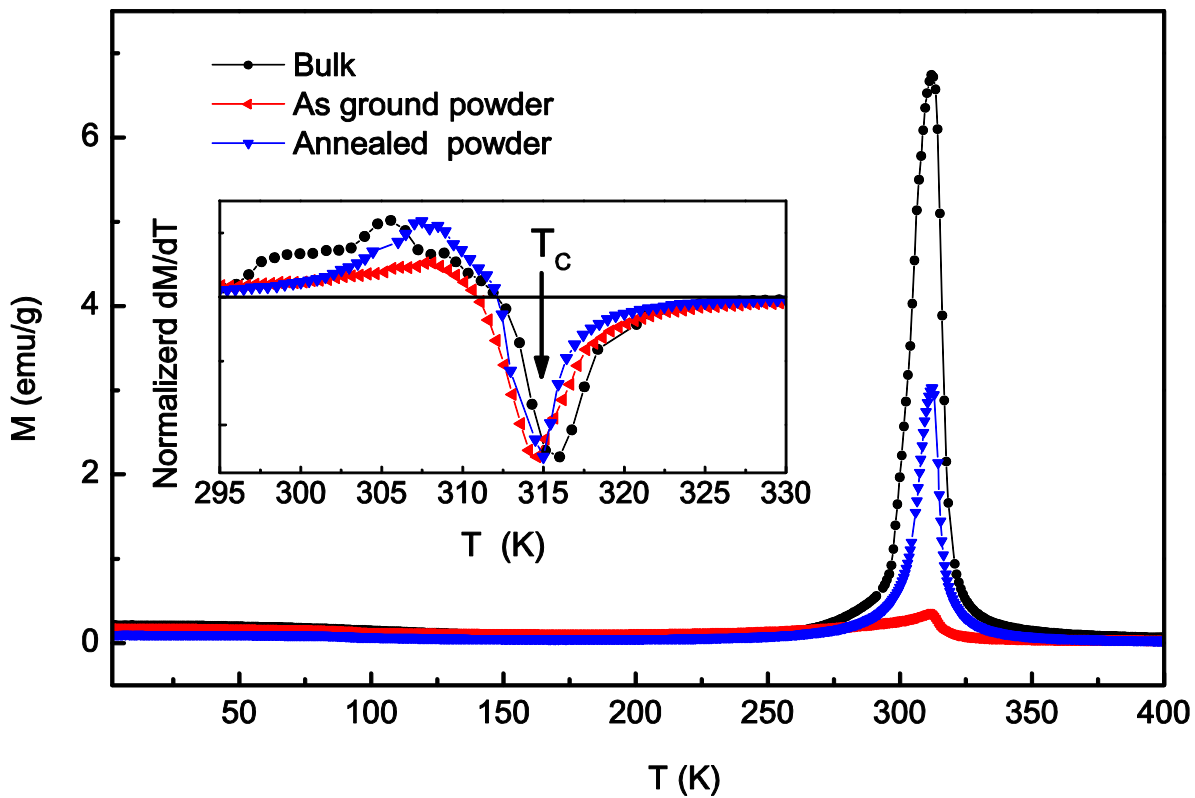

**Fig.5:** Magnetization as a function of temperature (a), and Rietveld fits for a few selected peaks for as-ground powder samples (b) and after annealing the same powder at 773 K for 10 hrs. (c). the labels I and II correspond to the results for $Mn_{1.75}Ni_{1.3}Ga$ and $Mn_{1.9}Ni_{1.1}Ga$ shown in (a), (b) and (c). The magnetization measurements involved a magnetic field of 50 Oe during cooling (FC) and warming (FW). The martensite start temperature $M_s$ is marked in (a). C and M represent peaks related to cubic austenite and martensite phases, respectively in (b) and (c).

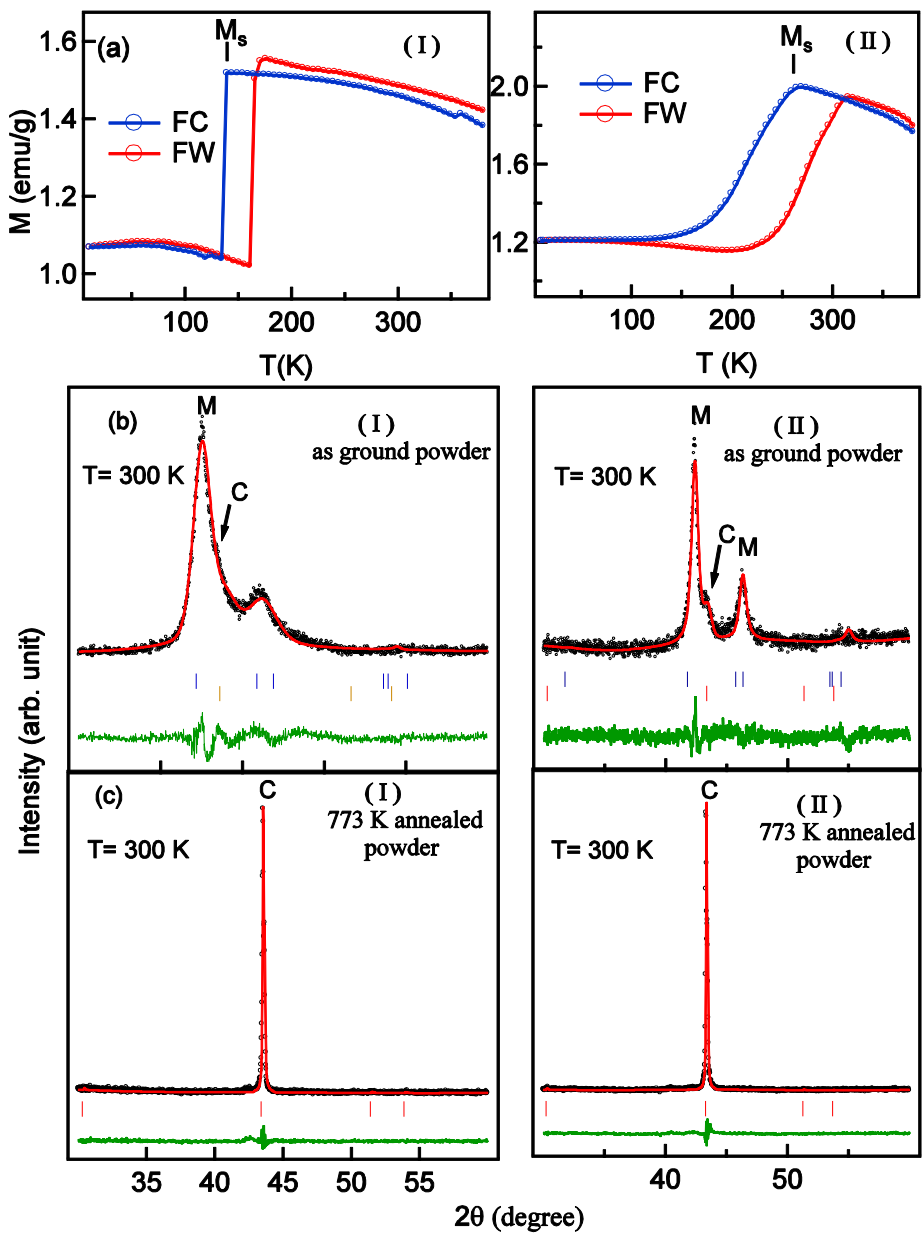

# Supplementary information

# Residual stress induced stabilization of martensite phase and its effect on the magneto-structural transition in Mn rich Ni-Mn-In/Ga magnetic shape memory alloys

Sanjay Singh[1, 2], Pallavi Kushwaha[2], F. Sceibel[1], Hanns-Peter Liermann[3], S. R. Barman[4], M. Acet[1], C. Felser[2] and Dhananjai Pandey[5]

[1]Experimentalphysik, University Duisburg-Essen, D-47048 Duisburg, Germany

[2] Max-Planck Institute for Chemical Physics of Solids, Nöthnitzer Strasse 40, 01187 Dresden, Germany

[3]Photon Sciences, FS-PE, Deutsches Elektronen Synchrotron (DESY), 22607 Hamburg, Germany

[4]UGC-DAE Consortium for Scientific Research, Khandwa Road, Indore, 452001, Madhya Pradesh, India.

[4]School of Materials Science and Technology, Indian Institute of Technology (Banaras Hindu University), Varanasi-221005, India.

**Experimental details:**

Polycrystalline ingots of $Mn_{1.8}Ni_{1.8}In_{0.4}$, $Mn_{1.75}Ni_{1.25}Ga$ and $Mn_{1.9}Ni_{1.1}Ga$ were prepared by melting appropriate quantities of the constituent metals of 99.99% purity under argon atmosphere using an arc furnace. This was followed by annealing the ingots in vacuum at 973 K ($Mn_{1.8}Ni_{1.8}In_{0.4}$) for one day and at 1100 K ($Mn_{1.75}Ni_{1.25}Ga$ and $Mn_{1.9}Ni_{1.1}Ga$) for nine days in sealed quartz ampules and quenching the annealed ingots into ice water. Pieces cut from the ingots were mechanically ground into powder using an agate mortar and pestle. In order to remove the stress introduced during grinding, the powders were annealed at 773 K under high vacuum [1, 2]. Both the annealed and as-ground powders were investigated by x-ray diffraction (XRD) and magnetization measurements. The XRD measurements were performed at the P02 beam line of Petra III, Hamburg, Germany at a wavelength of 0.20712 Å and also using a laboratory rotating Cu anode based powder diffractometer. The x-ray diffraction patterns were analyzed by Rietveld technique using Jana2006 software package. [3] The temperature dependence of the magnetization M (T) were measured using a superconducting quantum interference device magnetometer. The low field M (T) measurements were carried out using a small field of 100 Oe during warming after zero field cooling (ZFC), field cooling (FC) and field warming (FW).

Fig.1: X-ray diffraction patterns of $Mn_{1.8}Ni_{1.8}In$ at 340 K): (a) as-ground powder and (b) after vacuum annealing of the as-ground powder at 773K for 10 hrs. Inset in (b) shows retained martensite phase peaks (marked as M). The intensity of these peaks is ~0.9 % of the most intense peak and could be observed only in the high resolution synchrotron XRD patterns.

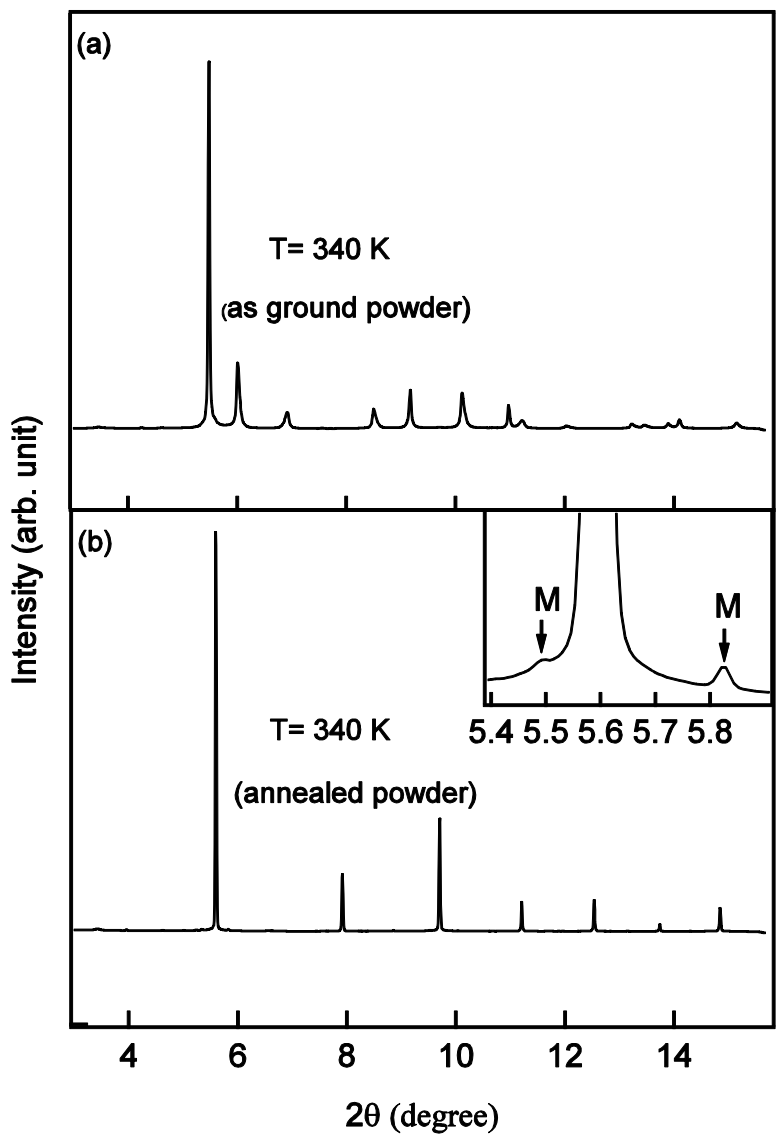

Fig. 2.   Williamson–Hall analysis of the XRD pattern of the as ground powder (Fig.1a) of Mn$_{1.8}$Ni$_{1.8}$In at 340 K.

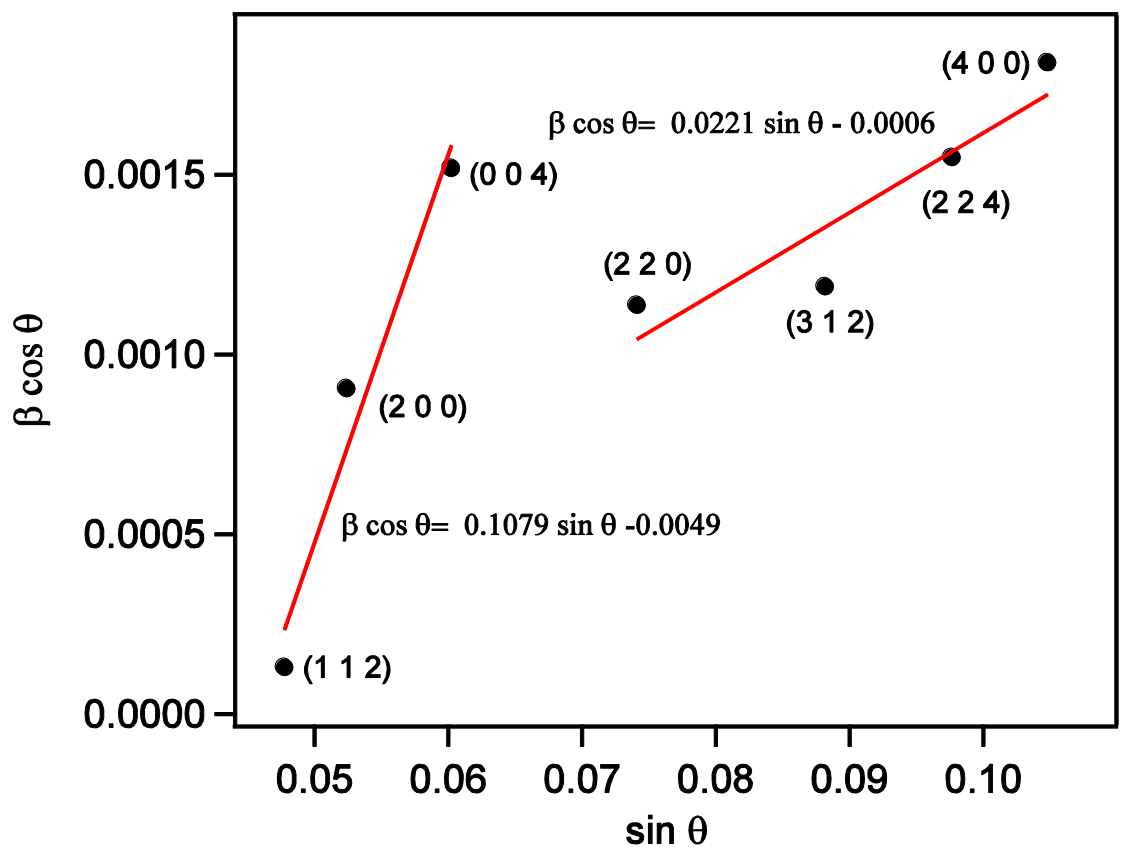

**Table. 1:** Unit cell parameters obtained from the Rietveld refinements and residual strain and coherently scattering domain size obtained from Williamson-Hall-analysis of the XRD patterns of the as-ground and annealed (773K) powders of $Mn_{1.8}Ni_{1.8}In_{0.4}$, $Mn_{1.75}Ni_{1.25}Ga$, $Mn_{1.9}Ni_{1.1}Ga$.

| Sample name and parameters | As ground powder | Annealed powder |
|---|---|---|
| **$Mn_{1.8}Ni_{1.8}In_{0.4}$** | | |
| Structure (T= 300 K): Lattice parameters: | Tetragonal (M) + Cubic<br>$a=b=3.9457$ (1)Å, + $a= 6.006$ Å<br>$c= 6.8757$ (3) Å | Cubic (A)<br>$a= 6.00482(2)$ Å |
| Phase fraction (%): | 97 + 3 | 99 |
| Volume ($Å^3$) | 107.044 | 216.52 |
| Microstrain ($\varepsilon$) | 5.4% | 0.02% |
| Domain Size (Å) | 194 | 550 |
| **$Mn_{1.75}Ni_{1.25}Ga$** | | |
| Structure (T= 300 K): Lattice parameters: | Tetragonal (M) + Cubic (A)<br>$a=b=3.93$Å, $a= 5.86$ Å<br>$c= 6.58$ (4) Å | Cubic (A)<br>$a= 5.8864(2)$ Å |
| Phase fraction (%): | 75 + 25 | 100 |
| Volume ($Å^3$) | 101.627 | 203.96 |
| Microstrain ($\varepsilon$) | 5.6% | 0.5% |
| Domain Size (Å) | 70 | 1025 |
| **$Mn_{1.9}Ni_{1.1}Ga$** | | |
| Structure (T= 300 K): Lattice parameters: | Tetragonal (M) + Cubic (A)<br>$a=b=3.920$ (1)Å, $a= 5.890(2)$ Å<br>$c= 6.678$ (3) Å | Cubic (A)<br>$a= 5.9009(1)$ Å |
| Phase fraction (%): | 90 + 10 | 100 |
| Volume ($Å^3$) | 102.61 | 205.38 |
| Microstrain ($\varepsilon$) | 3% | 0.35% |
| Domain Size (Å) | 110 | 768 |